\documentstyle[epsbox,12pt]{article}                               
\newcommand{\simgt}{\lower.5ex\hbox{$\; \buildrel > \over \sim \;$}}
\newcommand{\simlt}{\lower.5ex\hbox{$\; \buildrel < \over \sim \;$}}
\begin{document} 
\title{Specifying the Environments around GRB, Explaining the Fe Line in 
the X-Ray Afterglow of GRB000214 }
\author{{Kei \textsc{Kotake} and Shigehiro \textsc{Nagataki}}\\
{ \it{Department of Physics, The University of Tokyo,
7-3-1 Hongo, Bunkyo},}\\ { \it{Tokyo 113-0033, Japan}}\\
\it{E-mail: kkotake@utap.phys.s.u-tokyo.ac.jp}\\
\and
\it{E-mail: nagataki@utap.phys.s.u-tokyo.ac.jp}}

\maketitle 
\baselineskip=24pt
\begin{abstract}
 We present a model for explaining the Fe K$\alpha$ line and the continuum in the afterglow of  GRB 000214. In this paper, we pose the importance to seek the physically natural environment around GRB 000214. For reproducing the observation, we need a ring-like remnant around the progenitor, like that of SN 1987A produced by the mass-loss of the progenitor and the fireball spread over all directions. The observation of GRB 000214, in which the continuum power-law spectrum decreased faster than the line, motivated us to consider two independent systems for the line emission and the continuum spectrum. At first, the continuum spectrum can be fitted by the afterglow emission of the fireball pointing toward the observer, which does not collide with the ring because the emission of GRB and the afterglow are highly collimated to the observer by the relativistic beaming effect. Secondly, the line can be fitted by the fluorescence of the Fe atoms in the ring illuminated by the X-ray afterglow. 
The significance of this study is that our model may strongly constrain the GRB models. 

$KeyWords${ gamma-rays: bursts  ---  X-rays: general ---  ISM: supernova 
remnants  --- line: formation --- gamma-rays: individual (GRB 000214)}

\noindent

\end{abstract}


\section{Introduction} 

   There are four Gamma-Ray Bursts (hereafter GRBs) displaying the Fe K$\alpha$ emission lines in their afterglows (\cite{rf:Piro1},\cite{rf:Yoshida},\cite{rf:Piro2},\cite{rf:Antonelli}). What are the implications of these observations? A fireball model \cite{rf:Rees1}, which explains the behaviors of GRBs well, needs a central engine to give an initial energy input to the fireballs. What is a central engine? This has been a long-term mystery and a controversial
problem. However, with lines in the afterglows, we have an important clue for understanding the environments around GRBs. To construct a model which explains the line emission, the burst and the afterglow at the same time are of wide interest. Thus,  many models which had been proposed so far before the first line detection in 1997 should be modified to explain the line emission. Amongst them, we categorize them into two: binary-neutron-star merger models \cite{rf:Pacz1},\cite{rf:Eisher} and massive-star-related models (e.g., \cite{rf:Rees2},\cite{rf:Woosley},\cite{rf:Pacz2}). Between them, we take the massive-star-related models. This is because binary-neutron-star merger models, which produce GRBs during coalescence, take a time of almost the order of $10^9$ yr to merge and it happens far away from the star-forming regions. This contradicts with the observational facts that GRBs have often been detected in the 
star-forming region 
(e.g., \cite{rf:Boea},\cite{rf:Dar}). 
On the contrary, the following two facts strongly motivate us to consider the massive-star-related models: (1) that the possible association of GRB 980425 with the type Ic supernova 1998bw was observed (\cite{rf:Galama}), in spite of the chance probability for a spatial and temporal coincidence of GRB with the type Ic supernova being less than $10^{-4}$, and (2) that the detections of GRBs in the star-forming regions have been counted near twenty events, as stated above 
(e.g.,\cite{rf:Boea},\cite{rf:Dar}).  

 Now, which model should we take amongst massive-star-related models? Two well-known scenarios are supranova model (\cite{rf:Vietri1},\cite{rf:Vietri2}) and hypernova model (\cite{rf:Rees2},\cite{rf:Woosley},\cite{rf:Pacz2}). In the scenario of  supranova model, at first, the usual supernova explosion creates a rapidly spinning neutron star. Secondly, several months or years later, the neutron star, spinning down by emitting gravitational and electro-magnetic waves, collapses into a Kerr black hole. Thirdly, the black hole's rotational energy is converted to power the GRB \cite{rf:Blandford}. They assumed the ring remnants, produced during the supernova explosion. However it may not be probable, since the usual supernovae remnants have shell structures. In addition, if the morphology of the remnant around GRBs is shell-like, a fireball will be scattered inevitably by the supernova remnants to decelerate it within a day and will probably show detectable X-ray absorption or emission lines \cite{rf:Lazzati}. This contradicts the observation of X-ray afterglow of GRB970508 which lasted for a year and showed no detectable X-ray absorption or emission lines. In their latest paper  \cite{rf:Vietri2}, they assume a plerionic remnant, but it may be too speculative because of its special morphology. Among hypernova models  (\cite{rf:Rees2},\cite{rf:Woosley},\cite{rf:Pacz2}),  \cite{rf:Rees2} considered a situation, in which an extended magnetically-dominated wind from a GRB impacts the expanding envelope of a massive progenitor to produce the Fe line, and most of the continuum could still be explained from the standard decelerating fireball.  Also \cite{rf:Weth}  successfully reproduced the line and the continuum at the same time. Here, we suggest an alternative and natural model.
 We assumed a ring-like remnant, such that observed by HST around SN 1987A (e.g., \cite{rf:Plait},\cite{rf:Lundqvist}). In \cite{rf:Boea}, they discussed a strongly anisotropic GRB environment; that is, a  ring. They calculated excellently the time evolution of the interaction between the ring and the fireball, as well as various radiative processes, photoionization, fluorescence, recombination, electron-impact ionization, Compton scattering, bremsstrahlung and Coulomb scattering. As a result, they partially succeeded in reproducing the spectrum of  GRB 970508. According to their model, however, the continuum spectrum is dominated by the thermal bremsstrahlung emitted in the shocked region between the incoming fireball and the ring, which contradicts the power-law continuum suggested from the four above-mentioned observations with Fe lines. Also, in \cite{rf:Boeb}, matter amounting to $1M_\odot$ with a velocity of $\simeq 10^9$ cm s$^{-1}$ created by hypernova model \cite{rf:Fryer} is ejected along the equatorial plane to hit the ring created during the merger of the progenitors evolved to helium stars. However thermal bremsstrahlung was also dominant over the spectrum. If the systems of both the line-emission region and the continuum flux region are the same, the above problem seems to still be inevitable.  

 We present a model to reproduce the observation of GRB 000214, in which we consider two systems for the line emission and the continuum. That is to say, the  Fe K$\alpha$ line is produced by the fluorescence of the Fe atoms in the equatorial ring illuminated by the X-ray afterglow. On the other hand, the continuum power-law flux is obtained by the afterglow emitted from the fireball toward the observer (see. figure 1). Moreover, we considered the interaction between the ring and the fireball to investigate
whether the thermal bremsstrahlung emission from the shocked region, which was dominant for the continuum in \cite{rf:Boea}, is real or not. As a result, we find that the thermal bremsstrahlung emission from the shocked region is negligible by elaborately estimating the temperature behind the shock wave.

 In this paper, we propose a more clear picture to explain the observational facts; i.e., flux level, spectrum shaping and duration of the Fe line. Observational evidences of of GRB 000214 are described in section 2. In section 3, a physical picture for our model is stated. Discussion are presented in section 4.   
\section{Observations of GRB 000214} \label{extinction}  
 
 We give here the observation of GRB 000214 \cite{rf:Antonelli}. GRB 000214 had a fluence of $ F_{\gamma} = 1.4 \times 10^{-6}$ erg \rmfamily{cm}$^{-2}$  and a duration of $ t_{\gamma} \simeq 10 $ s in the energy band of the Gamma-Ray Burst Monitor on board the {\rm BeppoSAX} satellite (40 -- 700 keV). The fluence of the prompt X-ray was $1.0 \times 10^{-6}$ erg cm$^{-2}$ (2 -- 10 keV), which was detected by the Wide Field Cameras, also on board the {\rm BeppoSAX} satellite. A follow-up observation with the {\rm BeppoSAX} Narrow Field Instruments began about 12 hr after the corresponding GRB, and lasted for 104 ks. The effective exposure time was 51000s on source-time for {\rm BeppoSAX} Medium-Energy Concentrator/ Spectrometer (MECS) and 15000s in the Low-Energy Concentrator/ Spectrometer (LECS); the energy band for each was 1.6 -- 10 keV and 0.1 -- 4.0 keV. A spectral analysis using the data from MECS and LECS showed that the energy spectrum had a power-law $(\nu^{-\alpha})$ photon index of $(\alpha = 2.0 \pm 0.3)$ and $F_{\footnotesize{\mbox{x}}}$  (2 -- 10\,\,keV) = $(2.75 \pm 0.9) \times 10^{-13}$ erg cm$^{-2}$ s$^{-1}$. Line emission was identified with the Fe K$\alpha$ (for hydrogen-like iron) with a cosmological redshift of $\it{z}$ = 0.47. For the Fe line fitting in the spectrum, a narrow Gaussian was the best fit, which had a line-centroid energy of $E_{\footnotesize{\mbox{line}}}$ = 4.7 $\pm$ 0.2 keV and $F_{\rm{Fe}}$ =$ (1.00 - 3.02) \times 10^{-13}$ erg cm$^{-2}$ s$^{-1}$. The Gaussian fit for the emission line was too narrow to find the intrinsic velocity of the matter.  
For GRB 000214, from its narrowness, it can only be inferred that the intrinsic velocity was at most sub-relativistic.  
 We focus on the fact that the continuum flux decreased  faster than the
line flux during the observation of the X-ray afterglow. This motivates us to think about independent systems for the line-emitting region and the continuum-emitting region.    
\section{Physical Picture for Our Model}
\subsection{Motivations of Our Model} 
What is the motivation for our model? As written at the end of section 2, the continuum flux decreased faster than the line flux during the observation
of the X-ray afterglow. Also the observed continuum spectrum was power-law. If both the line emitting-region and the region for the continuum spectrum were supposed to be the same place, as assumed by \cite{rf:Boea}, the line would disappear within about one day. In addition, thermal bremsstrahlung emission is dominant. Both of them are not in the cases for the observation of the afterglow of GRB 000214. Here, we suggest an idea in which the systems for the line-emitting region and the continuum-emitting region are different. That is to say,  non-thermal continuum spectrum of GRB 000214 \cite{rf:Antonelli} can be explained dominantly by the afterglow toward the observer, and 
the emission of the Fe K$\alpha$ line can be explained by the fluorescence of the Fe atoms 
in the ring illuminated by the X-ray afterglow. In our model, 
 we needed an anisotropic environment around GRB following 
\cite{rf:Boea} (see also figure.1). In addition, we assumed an anisotropic energy deposition of the burst. An anisotropic energy deposition means that the energy per unit solid angle is emitted much stronger along the jet axis toward the observer than other regions. Here, we should clarify the configuration of the energy deposition. Basically, the fireball is emitted in every direction. To characterize the anisotropic energy distribution of a burst, we introduce two angles, $\theta$ in radian and $\delta \Omega$ in steradian, which are explained below. Radiation from relativistically moving matter is beamed in the direction of the motion to within an angle of $\theta = {\gamma}^{-1}$ radian, where $\gamma$ is the Lorentz factor of the relativistically moving matter. Also let $\delta\Omega$ be the angular size (in steradian) of the relativistically moving matter that emits a burst. To keep the total energy minimum, we set $\delta\Omega$ to be equal to $4\,\pi\, {\gamma}^{-2}$. We define $E_{\footnotesize{\mbox{jet}}}$, which is the energy emitted within $\delta \Omega \simeq 4\,\pi\,{\gamma}^{-2}$ str in the cone-like region, as stated above (see figure.1). As for the ring, $ E_{\footnotesize{\mbox{ring}}}$ is emitted within $\delta\Theta$ str, which is the covering angle (in steradian) of the ring region.  Also, for all regions except the cone-like and ring regions, there is no way to specify the outflow of  energy. This is because it cannot be observed by us due to the relativistic beaming effect stated above. For the region except for the cone-like and ring regions, $E_{\footnotesize{\mbox{other}}}$, which is assumed to be 
\begin{equation}
E_{\footnotesize{\mbox{other}}} = \frac{\delta E_{ring}}{\delta \Theta} \times(4 \pi - \delta \Theta - 2\,\delta \Omega),
\end{equation}
is emitted when we calculate the total energy of the system (see section 4). We note both $E_{\footnotesize{\mbox{jet}}}$ and $ E_{\footnotesize{\mbox{ring}}}$ represent the fireball's kinetic energy.  For clarity, we divide the physics involved here into two parts to explain the observation of GRB 000214. The fireball evolution toward us, which well explains the continuum spectrum, is in subsection 3.2, and the shock interaction of the fireball with the ring is in subsubsection 3.3.1, stating that the thermal bremsstrahlung emission from the shocked region is negligible. In subsubsection 3.3.2, the line emission mechanism is stated.
\subsection{Parametrization for the Continuum Afterglow}
 We analyze the afterglow of GRB 000214 following Sari, Piran and Narayan \cite{rf:Sari}.  Seven free parameters required to determine the spectrum in their model are determined below. $E_{\footnotesize{\mbox{spherical}}}$ is given as
\begin{equation}
 E_{\footnotesize{\mbox{spherical}}} = \frac{4\pi}{\delta \Omega}\,\,E_{\footnotesize{\mbox{jet}}} 
\end{equation}
(see figure.1), which is the kinetic energy of the fireball, estimated as if the explosion were spherical. In fact, if the explosion is jet-like, the intrinsic energy, $E_{\footnotesize{\mbox{jet}}}$, is much smaller. Also, $\epsilon_B$ is a parameter, which measures the ratio of the magnetic field energy density to the total thermal energy, $\epsilon_{\rm{e}}$ is a parameter, which measures the fraction of the total thermal energy which goes into the electrons in the thermal motions,  $n_{\footnotesize{\mbox{ism}}}$ is the number density for the interstellar matter, $t_{\rm{d}}$ is days from the burst, $\gamma_0$ is the initial fireball's Lorentz factor and $\it{D}$ is the distance from the GRB center to the
observer. Actually, from section 2, $t_{\rm{d}}$ = 0.5, $D = 5.8 \times 10^{27}$ cm, while $n_{\footnotesize{\mbox{ism}}}$ is set to 1 cm$^{-3}$. Then, the number of free parameters are reduced to four. We fix $E_{\footnotesize{\mbox{spherical}}} = 5.0 \times 10^{52}$ ergs, $\epsilon_B =\epsilon_{\rm{e}} = 0.5$ and $\gamma_{0} = 200$. We notice here that about one tenth of  $E_{\footnotesize{\mbox{spherical}}}$ is emitted as gamma-ray (e.g.,\cite{rf:Piran}).  The radiation of X-Ray afterglow of GRB000214 is specified as adiabatic and fast cooling with the above parameters (e.g., \cite{rf:Sari}). We explain it shortly below.
\begin{center}
\begin{tabular}
{|l||l|l|} \hline
   Types& Adiabatic Hydrodynamics & Radiative Hydrodynamics   \\ \hline\hline
   Slow Cooling & Arbitrary $\epsilon_{\rm{e}}$ & impossible  \\ \hline
  Fast Cooling & $\epsilon_{\rm{e}} \le 1$ &  $\epsilon_{\rm{e}} \simeq 1$\\ \hline
\end{tabular}
\end{center}
Table1 : Classification of the X-Ray afterglow emission (e.g.,\cite{rf:Piran} ) .\\

 The radiation of GRB afterglows is generally divided into three types (see Table 1). At first, we determine whether the radiation is fast or slow cooling. To do this,  we are required to prepare the two quantities, $\gamma_{\rm{c}}$, and $\gamma_{\rm{m}}$. Both of them are given like below: 
\begin{equation}
 \gamma_{\rm{c}} = \frac{3 m_{\rm{e}}}{16\,\epsilon_B \,\sigma_{\rm{T}}\, m_{\rm{p}}\, c\, t\, \gamma_{0}^3\,n_{\footnotesize{\mbox{ism}}}};
\end {equation}
If a single electron with Lorentz factor : $\gamma_{\rm{c}}$ loses its all kinetic energy by the synchrotron radiation, it means that the electrons with Lorentz factor greater than $\gamma_{\rm{c}}$ are rapidly cooled.  On the other hand, as in \cite{rf:Piran},
\begin{equation}
 \gamma_{\rm{m}} \simeq 610\,\epsilon_{\rm{e}}\,\gamma_{0}.
\end {equation}
Above $\gamma_{\rm{m}}$, the number density of electrons,$ N(\gamma)$,accelerated behind the shock obeys a power-law as $ N(\gamma) \propto{\gamma}^{-p} d\gamma,\,(\gamma \ge \gamma_{\rm{m}})$. If $\gamma_{\rm{c}} \leq \gamma_{\rm{m}} $, it is fast cooling, because the large amount of the electrons are rapidly cooled down by the synchrotron radiation. On the contrary, if  $\gamma_{\rm{m}} \leq \gamma_{\rm{c}} $, it is a slow cooling, only small fraction of the electrons can be cooled. In our case, it is the fast cooling with the above parameters. Next, we determine whether the radiation is radiative or adiabatic. To do this, we need the quantity $\epsilon_{\rm{e}}$ , which is the fraction of the total thermal energy $e$ which goes into the energy, $U_{\rm{electron}}$, of the electrons in the random motions, and given as 
\begin{equation}
\epsilon_{\rm{e}} = \frac{U_{\rm{electron}}}{e}.
\end{equation}
 As the exact value of $\epsilon_{\rm{e}}$ cannot be determined from the observation, it is a free parameter. We set $\epsilon_{\rm{e}} = 0.5$ to fit the observation.
In the end, we conclude the radiation is adiabatic and fast cooling. 
 For an adiabatic and fast cooling blast wave (from \cite{rf:Sari}),
\begin{eqnarray}
\nu_{\rm{c}}&=& 1.2 \times 10^{12}\,\,{\epsilon_B}^{-3/2}\,\Bigl(\frac{E_{\footnotesize{\mbox{spherical}}}}{5\times10^{52}\,\,\mbox{erg}}\Bigr)^{-1/2}\,\Bigl(\frac{n_{\footnotesize{\mbox{ism}}}}{1 \rm{cm}^{-3}}\Bigr)^{-1}\,t_{\rm{d}}^{-1/2}\,\,\mbox{Hz},\\
\nu_{\rm{m}}&=& 1.3 \times 10^{15}\,\,{\epsilon_B}^{1/2}\,{\epsilon_{\rm{e}}}^{2}
\Bigl(\frac{E_{\footnotesize{\mbox{spherical}}}}{5\times10^{52}\,\,\mbox{erg}}\Bigr)^{1/2}\,t_{\rm{d}}^{-3/2}\, \mbox{Hz},\\
F_{\nu,\rm{max}} &=& 1.7 \times 10^{6}\,{\epsilon_B}^{1/2}\,\Bigl(\frac{E_{\footnotesize{\mbox{spherical}}}}{5\times10^{52}\,\,\rm{erg}}\Bigr)\,
\Bigl(\frac{n_{\footnotesize{\mbox{ism}}}}{1 \rm{cm}^{-3}}\Bigr)^{1/2}\,\Bigl(\frac{D}{5.8\times10^{27}\rm{cm}}\Bigr)^{-2}\,\mu \mbox{Jy},
\end{eqnarray} 
where $\nu_{\rm{c}}$ and $\nu_{\rm{m}}$ correspond to the relevant Lorenz factors, $\gamma_{\rm{c}}$ and $\gamma_{\rm{m}}$, for the observed energy range of 2 -- 10 keV. The flux of the afterglow, $F_{\footnotesize{\mbox{x}}}$, can be estimated as
\begin{equation}
F_{\footnotesize{\mbox{x}}} = \Bigl(\frac{\nu_{\rm{m}}}{{\nu_{\rm{c}}}}\Bigr)^{-1/2}\,\Bigl(\frac{\nu}{\nu_{\rm{m}}}\Bigr)^{-p/2} F_{\nu,\rm{max}}\,\, \mbox{erg}\,\,\,\mbox{cm}^{-2}\,\mbox{s}^{-1}\,\rm{Hz}.
\end {equation}
, where we take the value of p to 4 to fit the observation.
In figure 2, we can fit the observed continuum spectrum well.  Near 1 keV, although it may seem that the spectrum is different, it is due to the absorption (Murakami, private communication). Thus, essentially, the spectrum is a good reproduction of the observation.    
\subsection{Parametrization of the Interaction between the Fireball and the Ring} 
 We divide this subsection into two to explain the Fe line from the shocked 
ring. At first, we state the thermal history of the ring, and  secondly the line-emission mechanism.  
\subsubsection{Thermal History of the Ring}
 We parameterize the ring. We assume a ring mass, $ M_{\footnotesize{\mbox{ring}}}$ of $ 9.0\times 10^{32}\,\,\mbox{g} $, an inner
radius, $R_{\footnotesize{\mbox{in}}}$ of $3.0\times 10^{15} \mbox{cm}$, the tenfold iron overabundance with respect to the solar abundance, $A_{\odot\,,\rm{Fe}}$ of $10$; the energy shedding
towards the ring, $E_{\footnotesize{\mbox{ring}}}$ is assumed to be $ 3.4 \times 10^{51}$ erg as the fireball's kinetic energy, the covering angle of $\phi = 50^{\circ}$ (see figure1), and the width of the ring $\delta$R to be  $3.0\times 10^{15}$ cm, which is from the duration of the observation lasting about 100 ks.  We notice here that about one tenth of $E_{\footnotesize{\mbox{ring}}}$ is emitted as gamma-ray (e.g., \cite{rf:Piran}) and about one thousandth of $E_{\footnotesize{\mbox{ring}}}$ is emitted as X-ray, it is inferred by comparing the fluence of the gamma-ray with that of X-ray in the observation of GRB 000214 (e.g., \cite{rf:Antonelli}). From the observation, $R_{\footnotesize{\mbox{in}}}$ is determined by the time lag by about one day time lag between the occurrence of GRB and the appearance of the Fe line,
\begin{equation}
 R_{\footnotesize{\mbox{in}}} = c\,\,\frac{\mbox{1\,\,day}}{T_{\footnotesize{\mbox{time-lag}}}} = 3\times 10^{15}\,\,\mbox{cm}.
\end {equation} 
 The fireball emits GRB at $ R_{\gamma} \simeq 10^{12}$ cm and  
the afterglow at $R_{\footnotesize{\mbox{afterglow}}} \simeq 10^{14}-10^{15}$ cm, provided that one hundredth of the number density of the ring prevails to $R \simeq 10^{14}$ cm (e.g., \cite{rf:Sari}).
  
 After the half day since the ignition of the fireball, powered by the explosion energy $E_{\footnotesize{\mbox{ring}}}$, the fireball with the initially loaded mass of $M_{0}$ hits the ring. If we define $R_{\footnotesize{\mbox{d}}}$, where the fireball sweeps the amount of  mass ($\simeq \frac{M_{0}}{\gamma_{0}}$), then  $ t_{\footnotesize{\mbox{sub-rela}}}$  after the interaction the blast wave will be decelerated to sub-relativistic speed (e.g., \cite{rf:Mckee}):
\begin{equation}
 R_{\rm{d}} =\Bigl( \frac{3\,E_{\footnotesize{\mbox{ring}}}}{4\pi\,\cos\phi\,n_{\footnotesize{\mbox{ring}}}\,m_{\rm{p}}\,c^2\,{\gamma_0}^2} + R_{\footnotesize{\mbox{in}}}^3\Bigl)^{1/3} \simeq 10^{15}\,\,\, \mbox{cm}
\end {equation}
and 
\begin{equation}
 t_{\footnotesize{\mbox{sub-rela}}} \simeq \frac{R_{\footnotesize{\mbox{d}}} - R_{\footnotesize{\mbox{in}}}}{c}  = 1.2 \times 10^{-2}\,\,\,\mbox{s}.
\end {equation}
It shows that before $t_{\footnotesize{\mbox{sub-rela}}}$, the ultra-relativistic behavior of the shocked fluids can be described by the Blandford-McKee solution \cite{rf:Kobayashi}, and after $t_{\footnotesize{\mbox{sub-rela}}}$ it can be described by the Sedov-Taylor solution. We should estimate the temperature in the two regions, that is to say, that of the ultra-relativistic region (hereafter URR), and of the Newtonian region (hereafter NR).  
At first, for URR, the energy density and the number density of the shocked region are estimated analytically ( e.g., \cite{rf:Piran})
\begin{equation}
   e(r,t) = 4\,n_{\footnotesize{\mbox{ring}}}\,\,m_{\rm{p}}\,\,c^2\,\, \gamma(t)^2\{1 + 16 \gamma(t)^2\,\,(1 - r/R)\}^{-17/12}\,,
\end{equation}
\begin{equation}
   n(r,t) = 4\,n_{\footnotesize{\mbox{ring}}} \gamma(t)\{1 + 16 \gamma(t)^2\,\,(1 - r/R)\}^{-5/4}\,.
\end{equation}
where $\it{R}$ is the distance of the shock wave measured from $R_{\footnotesize{\mbox{in}}}$, $\it{r}$ is the distance of the matter behind the shock wave measured from $R_{\footnotesize{\mbox{in}}}$, and $\gamma(t)$ is the Lorentz factor of the shock wave in the rest frame of the unshocked ring. Both of them can be estimated from  $E_{\footnotesize{\mbox{ring}}}$ and $ n_{\footnotesize{\mbox{ring}}}$, as follows: (e.g., \cite{rf:Piran})
\begin{equation}
\gamma(t) = \frac{1}{4}\Bigr(\frac{17\,E_{\footnotesize{\mbox{ring}}}}{\pi\,\,
n_{\footnotesize{\mbox{ring}}}\,\,m_{\rm{p}}\,\,c^5\,\,t^3\,\,\cos\phi}\Bigl)^{1/8},
\end{equation}
\begin{equation}
R(t) = \Bigr(\frac{17 E_{\footnotesize{\mbox{ring}}} t}{\pi m_{\rm{p}} n_{\footnotesize{\mbox{ring}}}\,\,c\,\,\cos\phi}\Bigl)^{1/4}.
\end{equation}
 On the other hand, the energy density of the nucleus in the shocked region at a given temperature is 
\begin{equation}
 e(r,t) = \frac{3}{2}\,\,n(r,t)\,k_{B}\,\,T_{\footnotesize{\mbox{shocked ring}}}.
\end{equation}
Equating equations (13) with (17), we can estimate the upper limit of the temperature in the shocked region by setting  $ r = R$.
For the UUR, 
\begin{eqnarray}
T_{\footnotesize{\mbox{ URR}}} &=& 2.0\times 10^{12}
\Bigl(\frac{E_{\footnotesize{\mbox{ring}}}}{3.4\times10^{51}\,\,\rm{ergs}}\Bigr)^{1/8}
\,\,\Bigl(\frac{n_{\rm{ring}}}{1.4 \times 10^9\,\,\rm{cm}^{-3}}\Bigr)^{-1/8}\nonumber
\\
& & \times\Bigl(\frac{t}{8\times 10^4\rm{s}}\Bigr)^{-3/8}\Bigl(\frac{\sec\phi}{\sec50^{\circ}}\Bigr)^{1/8}\,\,\,\rm{K}.
\end{eqnarray}
Also, for the NR,
\begin{eqnarray}
T_{\footnotesize{\mbox{ NR}}}&=& 1.0\,\,\times 10^{12}  \Bigl(\frac{E_{\footnotesize{\mbox{ring}}}}{3.4\times10^{51}\rm{ergs}}\Bigr)^{2/5}\Bigl(\frac{n_{\rm{ring}}}{1.4\times10^9 \rm{cm}^{-3}}\Bigr)^{-2/5}\times \nonumber
\\ & & \Bigl(\frac{\sec\phi}{\sec50^{\circ}}\Bigr)^{2/5}\Bigl(\frac{t}{8\times 10^4\rm{s}}\Bigr)^{-6/5}\,\,\,\rm{K}.
\end{eqnarray}
In \cite{rf:Boea}, the temperature of the shocked region was estimated to be $10^8\,\,\rm{K}$, assuming that the total kinetic energy of the incoming fireball was roughly equal to the total thermal energy in the shocked region. However, for the explicit description of the thermal history of the ring, it is inevitable to deal with the shock. Therefore, we have solved the shock dynamics analytically to estimate the temperatures of shocked regions, which are classified into two cites, that is to say, URR and NR. 
 For the thermal bremsstrahlung emission in the shocked region, the intrinsic emissivity is given as
\cite{rf:Lightman}
\begin{eqnarray}
 {\varepsilon_{\nu}}^{ff} &=& 5.1 \times 10^{-26}\sum_{k=1}^{26}\Bigl(\Bigl(\,\frac{\sum_{j=1}^{26}Z_{j}\,X_{j}\,n_{\footnotesize{\mbox{shocked ring}}}}{A_{j}}\Bigr)\times\nonumber\\
& & \frac{{Z_{k}}^2\,X_{k}}{A_{k}}\,n_{\footnotesize{\mbox{shocked ring, k}}}\Bigr)\,\,\,T_{\footnotesize{\mbox{NR}}}^{-1/2}\times\nonumber\\
& &  \exp\Bigr({\frac{-h\,\nu}{k_{\rm{B}}\,T_{\footnotesize{\mbox{NR}}}}}\Bigl)\,\mbox{{erg}}\,\,\,\mbox{cm}^{-3}\,\,\mbox{s}^{-1}\,\,\mbox{Hz}^{-1},
\end{eqnarray}
, where the summation are taken from j=1 (hydrogen) to j=26 (iron), and $Z_{j},
X_{j}, n_{\footnotesize{\mbox{shocked ring,j}}}$ represent electron number,  mass fraction, the number density of the j species of nucleus, and the gaunt factor is excluded in the equation (20), which is on the order of unity (e.g., 
\cite{rf:Lightman}, page 161). We remark that both $n_{\rm{shocked\,\,ring}}$, and $T_{\rm{shocked\,\,ring}}$ in equation (20) are the quantities estimated in the NR region,  because after $t_{\rm{sub-rela}}$, the dynamics of the shock wave can be described by the Sedov-Taylor solution.
For the temperature in the equations (18) and (19), the thermal bremsstrahlung emission from the shocked region is negligible.  It is because under the relevant temperature, the energetic electrons cannot be bent by the Coulomb interaction with the nucleus, leading the emissivity of the thermal bremsstrahlung lower. And the fluorescent line emission will be suppressed, since all the nucleon will be ionized under the temperature for the production the fluorescent line. Therefore if the line-emitting region and the continuum-emitting region are the same,  as in \cite{rf:Boea}, neither the continuum nor the line can be explained. In our model, the problem can be resolved to take the two systems for the line-emitting region and the-continuum emitting region into consideration.
In our model, the continuum can be determined only by the afterglow pointing toward us (see figure 2).
\subsubsection{Line Emission Mechanism}
 For Fe K$\alpha$ emission, three mechanisms are proposed by \cite{rf:Lazzati}, namely,  fluorescence in an optically thin ring, thermal emission from the ring and reflection (see figure1 in \cite{rf:Lazzati}).
In our model, the Fe K$\alpha$ line is produced by the fluorescence scenario, because the others have problems. In a model of thermal emission from the ring, if the ring is thermalized by the shocked region to produce line emission, it contradicts with the observation as stated in subsection 3.1. In the reflection model, the line is explained by the fluorescent Fe line near the surface of the ring, if the ring is optically thick to the photons of the X-ray afterglow. However, just after the arrival time of the burst photon, the fireball will hit the ring and will alter the non-thermal plasma for the fluorescent line into the thermal plasma. It seems difficult to explain the line by the reflection model. For GRB 000214, to avoid this problem, we considered slightly ionized Fe atoms in an optically thin plasma. We emphasize here that the fluorescence comes from an optically thin region. As stated in subsection 3.3.1, after $ t_{\footnotesize{\mbox{sub-rela}}} \simeq 1\times 10^{-2}$ s the shock wave becomes non-relativistic and the dynamics of the shock can be described by the Sedov Taylor solution. Therefore, the radii of the shocked wave, which can reach after one day from the burst, can be estimated as,
\begin{equation}
R_{\footnotesize{\rm{shock}}} = R_{\rm{in}} + R_{\rm{N,R.\,\,one\,\,day}} = 4.7  \times 10^{15}\,\,\,\rm{cm}.
\end{equation}
  Thus, the effective volume ,${V_{\rm{emitting\,\,region}} = 1.9\times 10^{47}\,\,\rm{cm^{3}}}$ for the fluorescence is about half of the total volume of the ring, on the other hand, the volume of the shocked region is ${V_{\rm{shocked\,\,region}} = 2.1\times 10^{47}\,\,\rm{cm^{3}}}$ . The shock wave has passed half of the total volume of the ring, in which the plasma is thermalized so as not to emit the line by fluorescence. The time evolution of the fluorescence is stated roughly in the following. At first, photons from the X-ray afterglow are absorbed in the ring to render the electrons in the neutral Fe atoms ionized. Secondly, the fluorescent K$\alpha$ line is emitted by the transitions of the electrons. The process of making the fluorescence line is effective at the time when the ring is illuminated not by a prompt burst, but by the X-ray afterglow. This is because the recombination time-scale for explaining the observed line should be less than $1\times 10^{-7}\,\,\rm{s}$ if the line is produced by the prompt gamma-ray; on the other hand; the recombination time-scale during the prompt gamma-ray is less than $\simeq100$s (e.g.,\cite{rf:Lazzati}). We thus find that the line cannot be produced by the prompt gamma-ray. As we state in the following in detail, the recombination time required for the line is not very short for the X-ray afterglow.  To explain the line flux by fluorescence, we demand two conditions. The first one is that the plasma in the ring should not be fully ionized, because if the region is fully ionized, line should not be emitted. The second one is that the optical depth for the line should be on the order of unity. This is because if the region is optically thick to line photons, we cannot observe the line. In the following, we refer to the above two conditions. The first one is that the plasma in the ring illuminated by the X-ray afterglow should not fully ionize the plasma. For determining the ionization state of the plasma, which is illuminated by the incoming photons, we have only to compare the ionization time scale, $t_{\rm{ion}}$, with $t_{\rm{rec}}$. For this purpose, we should specify the temperature of the ring after illumination by the X-ray afterglow. The heating rate, $\Gamma$, can be estimated as (Nakayama, Masai in preparation),
\begin{eqnarray}
\Gamma & = & \frac{n_{\rm{Fe,\,\,K}}}{4\,\pi R_{\rm{in}}^2 \cos{\phi}}\int_{\epsilon_{\rm{{Fe,\,\,K}}}}^{\infty}\,\sigma_{\rm{{Fe,\,\,K}}}(\epsilon -\epsilon_{\rm{{Fe,\,\,K}}})\,\,\frac{dL}{d\epsilon}\,\,\frac{d\epsilon}{\epsilon}\nonumber\\
& = &  2.4\times 10^{-2}\,\,
\Bigl(\frac{R_{\rm{in}}}{3\times10^{15}\mbox{cm}}\Bigr)^{-2}\,
\Bigl(\frac{n_{\footnotesize{\mbox{ring}}}}{1.4\times 10^{9} \rm{cm}^{-3}}\Bigr)
\,\,
\Bigl(\frac{E_{\footnotesize{\mbox{ring}}}}{3.4\times10^{51}\,\,\mbox{erg}}\Bigr) 
\Bigl(\frac{A_{\odot, \rm{Fe}}}{10}\Bigl)\,(\frac{\sec{\phi}}{\sec{50^{\circ}}}\Bigr)\,\,\nonumber\\
& &\mbox{erg}\,\,\mbox{cm}^{-3}\,\,\mbox{s}^{-1}.
\end{eqnarray}
, where $\sigma_{\rm{{Fe, K}}}$ is the approximated photoionization cross section \cite{rf:BoeP},
\begin{equation}
\sigma_{\rm{{Fe,\,\,K}}} = 9.3\times10^{-21}\Bigl(\frac{\epsilon}{\epsilon_{\rm{k}}}\Bigr)^{-3}\,\,{\rm{cm}^{2}}
\end{equation}
, $\epsilon_{\rm{{Fe,\,\,K}}}$ is the edge energy of Fe K of 7.5 keV, and $\it{L}$ is the luminosity of the X-ray afterglow (from 1 keV), where we assume the spectrum obeys power--law, like that of the X-ray afterglow pointing toward us.
The quantity of  $t_{\rm{ill}}$, which is the illumination time by the X-ray afterglow of the ring, can be estimated as,
\begin{equation}
t_{\rm{ill}} = \frac{R {\rm in}}{2\,\,\gamma_{0}^2\,\, c}= 1.3\,\,\Bigl(
\frac{R_{\rm in}}{3\times10^{15}\,\,\rm{cm}}\Bigr)\Bigl(\frac{\gamma_{0}}{200}\Bigr)^{-2}\,\,\rm{s}.
\end{equation}
 The temperature of the illuminated ring of $T_{\footnotesize{\mbox{illuminated ring}}}$ can be estimated as,
\begin{equation}
\Gamma\,\,t_{\rm{ill}} = \frac{3}{2}\,\,n_{\rm{ring}}\,\,k_{\rm B}\,\,T_{\rm illuminated ring}\,\,\rm{erg}\,\,\rm{cm}^{-3}.
\end{equation}
and, 
\begin{eqnarray}
T_{\footnotesize\mbox{illuminated\,ring}} &=& 3.3\times 10^5\,
{\Bigl(\frac{R_{\rm{in}}}{3\times10^{15}\mbox{cm}}\Bigr)}^{-2} {\Bigl(\frac{E_{\rm{ring}}}{3.4\times10^{51}\,\,\mbox{erg}}\Bigr)}
\Bigl(\frac{\sec{\phi}}{\sec{50^{\circ}}}\Bigr)\Bigl(\frac{A_{\odot,\rm{Fe}}}{10}\Bigl)\nonumber\\
& &\,\,\mbox{K}.
\end{eqnarray}
 We can estimate the recombination time-scale for the Fe K$\alpha$ line as \cite{rf:Verner},
\begin{equation}
t_{\footnotesize{\mbox{rec}}} = 1.0\times10^{-3}\,\,{\Bigl(\frac{n_{\mbox{ring}}}{1.4\times10^{9}\rm{cm}^{-3}}\Bigr)}^{-1} \frac{1}{\alpha(\lambda)}\,\,\,\rm{s}.
\end{equation}
, where 
\begin{eqnarray}
\alpha(\lambda) &=& 1.4\times10^{-12}\,\,{\lambda}^{1/2}\,(0.5\ln\lambda + 4.2\times10^{-1} + \frac{5.0\times10^{-1}}{\lambda^{1/3}}) \,\,\rm{cm^{-3}\,\,s^{-1}}, 
\end{eqnarray}
and where $\lambda$ is given as,
\begin{eqnarray}
\lambda &=&  3.2\times10^2\,\,{\Bigl(\frac{R_{\rm{in}}}{3\times10^{15}\mbox{cm}}\Bigr)}^{2} \Bigl(\frac{\sec{\phi}}{\sec{50^{\circ}}}\Bigr)^{-2}\Bigl(\frac{A_{\odot,\rm{Fe}}}{10}\Bigl)^{-1}
{\Bigl(\frac{E_{\rm{ring}}}{3.4\times10^{51}\,\,\mbox{erg}}\Bigr)^{-1}}.
\end{eqnarray}
On the other hand, $t_{\footnotesize{\mbox{ion}}}$ can be estimated as (Nakayama, Masai in preparation)   
\begin{eqnarray}
t_{\footnotesize{\rm{ion}}} &=& \Biggl(\frac{1}{4\,\pi R_{\rm{in}}^2 \cos{\phi}}\int_{\epsilon_{\rm{{Fe,K}}}}^{\infty}\,\sigma_{\rm{{Fe,K}}}\,\,\frac{dL}{d\epsilon}\,\,\frac{d\epsilon}{\epsilon}\Biggr)^{-1}\nonumber\\
& = & 2.2\times10^{-3}\,\,{\Bigl(\frac{R_{\rm{in}}}{3\times10^{15}\mbox{cm}}\Bigr)}^{2}{\Bigl(\frac{E_{\rm{ring}}}{3.4\times10^{51}\,\,\mbox{erg}}\Bigr)}^{-1}\,\,\mbox{s}.
\end{eqnarray}
 With the above two time-scale, we can find that the time-scales for the ionization is comparable with that of recombination. In other words, the plasma is not fully ionized by the incident photons of the X-ray afterglow. We conclude that the fluorescence line is emitted from the non-thermal plasma. Secondly, we demand both the Thomson optical depth, $\tau_{\rm{T}}$, and the optical depth of the bound--free transition $\tau_{\rm{bf}}$ in the plasma be on the order unity. This is because the large optical depth broadens and vanishes the line feature. The optical depth of the Thomson scattering can be estimated as, 
\begin{equation}
\tau_{\rm{T}} = 6.0\times 10^{-1}\,\,\Bigl(\frac{n_{\footnotesize{\mbox{ring}}}}{1.4\times 10^9\,\,\rm{cm}^{-3}}\Bigr)\,\,\Bigl(\frac{\delta R_{\rm{emitting\,\,region}}}{9\times 10^{14}\,\,\rm{cm}}\Bigr). 
\end{equation}
The optical depth for the bound--free transition of a single Fe atom can estimated as (e.g.,\, \cite{rf:Bowers}), 
\begin{equation}
\tau_{\rm{bf}} =  5 \times 10^{-4}\Bigl(\frac{n_{\rm{ring}}}{1.4\times10^{9} 
\rm{cm}^{-3}}
\Bigr)\,\,\Bigl(\frac{\delta R_{\rm{emitting\,\,region}}}{9\times 10^{14}\rm{cm}}\Bigr)\Bigl(\frac{A_{\odot,\rm{Fe}}}{10}\Bigr).
\end{equation}
  Line photons are optically thin to the Thomson scattering and the bound--free transition in the illuminated ring and can escape from the ring without smearing too much the iron line. Hence, we can roughly estimate the line emission by fluorescence. 
From \cite{rf:Lazzati},    
\begin{eqnarray}
F_{\footnotesize{\mbox{Fe}}} &=& 1.3 \times 10^{-13}
\Bigl(\frac{n_{\footnotesize{\mbox{ring}}}}{1.4\times 10^{9}\,\, \rm{cm}^{-3}}\Bigr)\Bigl(\frac{q}{0.7}\Bigr)\Bigl(\frac{E_{\rm{ring}}}{3.4\times 10^{51}\,\,\mbox{erg}}\Bigr)\Bigl(\frac{A_{\odot,\rm{Fe}}}{10}\Bigr)\times \nonumber \\
& & {\Bigl(\frac{t}{8\times 10^4 \,\,\mbox{s}}\Bigr)}^{-1}\,\Bigl(\frac{R_{\footnotesize{\mbox{in}}}}{3\times 10^{15}\,\,\mbox{cm}}\Bigr)\Biggl(\frac{\frac{V_{\rm{emitting\, region}}}{1.9\times10^{47}\,\,\rm{cm^3}}}
{\frac{V_{\rm{emitting\, region}}}{1.9\times10^{47}\,\,\rm{cm^3}} +\frac{V_{\rm{shocked\, region}}}{2.1\times10^{47}\rm{cm^3}}}\Biggr)\, \,\mbox{erg}\,\,\,\,\mbox{cm}^{-2}\,\,\mbox{s}^{-1}
\end{eqnarray}
, where $\it{q}$ is the fraction of the total X-Ray afterglow fluence absorbed by the ring and reprocessed into the line; $\it{q}$ can be estimated as (e.g., \cite{rf:Ghisellini}) 
\begin{equation}
 q  = 0.7\,\,\,\,{\Bigl(\frac{R_{\rm{in}}}{3\times10^{15}\mbox{cm}}\Bigr)}{\Bigl(\frac{{n_{\footnotesize\mbox{{ring}}}}}{1.4\times10^9\,\,\rm{cm}^{-3}}\Bigr)}\Bigl(\frac{\frac{\epsilon_{\footnotesize{\mbox{edge}}}}{\epsilon_{\footnotesize\mbox{max}}}}{0.2}\Bigr)\\,
\end{equation}
 where  $\epsilon_{\rm{max}}$ is the maximum energy of the X-ray afterglow, which is on the order of several ten keV and  $\epsilon_{\rm{edge}}$ is the edge energy, which is 7.5 keV.
As a result, we reproduced the line (see figure.2).

\section{Discussion}
  We discuss the significance of this study concerning the mechanism that produces a GRB. No simulations, challenging to explain the stellar collapses and the birth of the GRB at the same time, have succeeded. In numerical simulations of collapsars (e.g., \cite{rf:Woosley}), a very steep collimation of the jet whose opening angle of about 1$^{\circ}$ is generated. On the contrary, from the observation of SN1998bw, only 1\% of the polarization in the optical band was detected, which showed that the explosion may not be so collimated.  In our model, the fireball should spread out in every direction for illuminating the Fe atoms in the equatorial ring to emit the Fe K$\alpha$ line by fluorescence, which may be consistent with the observation. This picture will strongly constrain the GRB models. 

 We also assumed in this study the existence of a ring which has a tenfold Fe  overabundance with respect to the solar abundance. You may have thought that this value is relatively high for an object whose redshift (z) is 0.5.
However, because the GRB are born in the star-forming region, the composition of a slightly high metalicity of Fe might be justified.

 It is noted that the optical flash of GRB 990123 \cite{rf:Reem} may be explained by the black--body emission from the shocked region of the ring. We will discuss this problem in the forthcoming paper.

 We refer to the amount of the total energy emitted in our system as the kinetic energy of a fireball. If we take the initial Lorentz factor as 200, then from equation (1),  $E_{\footnotesize{\mbox{jet}}}= 1.25\times 10^{48}$ erg within $3.14\times 10^{-4}$ str, and we can roughly estimate $5.3\times 10^{51}$  erg for the region except for the cone region. As a result, the total amount of kinetic energy is nearly $ 10^{52}$ erg. This energy is compatible with the usual GRBs.
 
   You may wonder why there are only four gamma-ray bursts displaying Fe line in the X-ray afterglow and the others are not. We present two explanations for this question. At first, this may be because of the variations of distance from an origin of gamma-ray burst to a ring. If the distance is too far away, the flux emitted from the fireball will be so small that the resulting intensity of
the Fe line is too weak to be observed. Secondly, this may be the variations of the opening angle of the jet. If this is so collimated as not to illuminate the ring, the line will not be emitted.

 It is concluded in our model that the angular size of the relativistically
moving matter that emits the burst should be very large (at least, the opening
angle should be larger than several degrees) in order to explain the
intensity of the iron line. Thus, there should be correlation
between the opening angle of the burst and existence of the iron line
emission in the X-ray afterglow. We hope that this tendency will be confirmed
by the observations in the near future.

 The duration of the line emission in the X-ray afterglow is nearly one day for GRB 970508 \cite{rf:Piro1}, on the other hand, is more than one day for GRB 000214 \cite{rf:Antonelli}. Why are there variations of the duration of the Fe line emission?  We consider this may be because of the variations of the width of a ring. If it is narrow, the line will disappear soon, on the other hand, if it is wide, the line will survive longer. We think the former is the case of GRB 970508, and the latter is that of GRB 000214.  

 It is a difficult problem to determine observationally whether the line photons originate from helium-like (6.4 keV) or hydrogen-like (6.95 keV) Fe atoms. For GRB 970828, it was determined that the latter is correct by detecting the emission lines from a possible host galaxy (Yoshida et al. in preparation), although there is no observation such like that in the case of GRB 000214.  
The ionization parameter in our model can be estimated as in
 \cite{rf:Piro1},
\begin{eqnarray}
\xi &=& \frac{L}{{n_{\rm ring}} R_{\rm in}^2} \nonumber\\
    &\simeq& 1\times 10^8\, \Bigl(\frac{E_{\rm ring}}{3.4\,\times\,10^{51}\,{\rm erg}}\Bigl)\Bigl(\frac{t_{\rm ill}}{1.3\,\rm{s}}\Bigr)^{-1}\Bigl(\frac{n_{\rm ring}}{1.4\,\times 10^9\,\,\rm{cm}^{-3}}\Bigr)^{-1}\Bigl(\frac{R_{\rm in}}{3\,\times\,\,10^{15} \rm{cm}}\Bigr)^{-2}\nonumber\\
& & \,\rm{erg\,\,cm}\,\,{\rm{s}}^{-1}.
\end{eqnarray}
 The Fe atoms are ionized for this value of the ionization parameter (e.g.,  \cite{rf:Piro1}). This indicates the possibility that the line originates from hydrogen-like Fe atoms in our model. To discuss whether the hydrogen-like or helium-like lines are prominent, we are supposed to require more detailed analysis on the emission-line features. Therefore, as a future work, we will perform a series of precise calculations on the ionization states in a ring illuminated by the X-ray afterglow and investigate the effects of fluorescence on the line emission in detail. 
 \section*{Acknowledgments}
The authors are grateful to Professor K. Masai, Professor H. Murakami and Dr. D. Yonetoku for useful discussions.

\newpage
\begin{figure}[h]
\begin{center}
\psbox{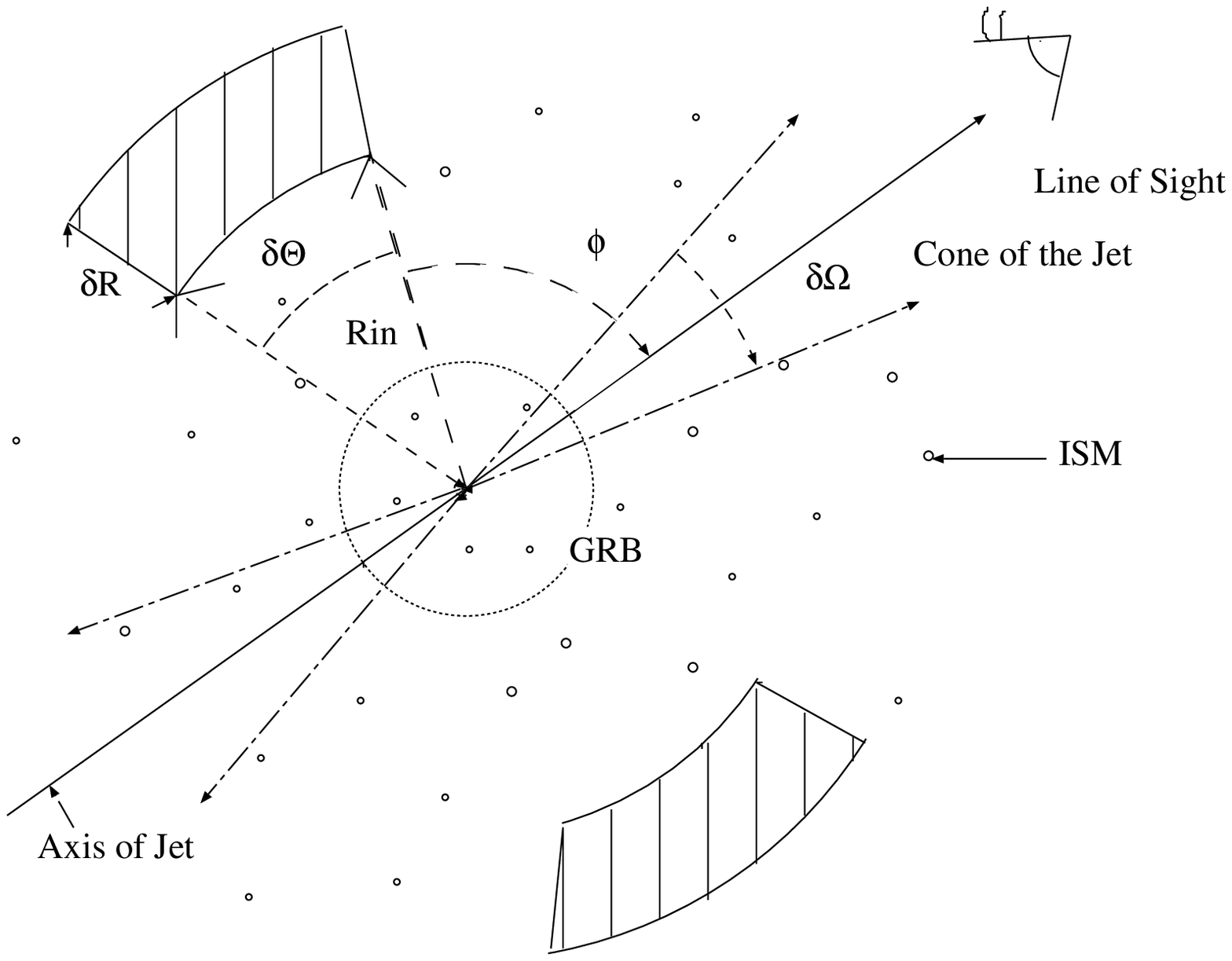}
 \caption{Schematic diagram of the GRB's environment in our model.}
 \label{Fig:1}
\end{center}
\end{figure}

\newpage
\begin{figure}[h]
\begin{center}
\psbox{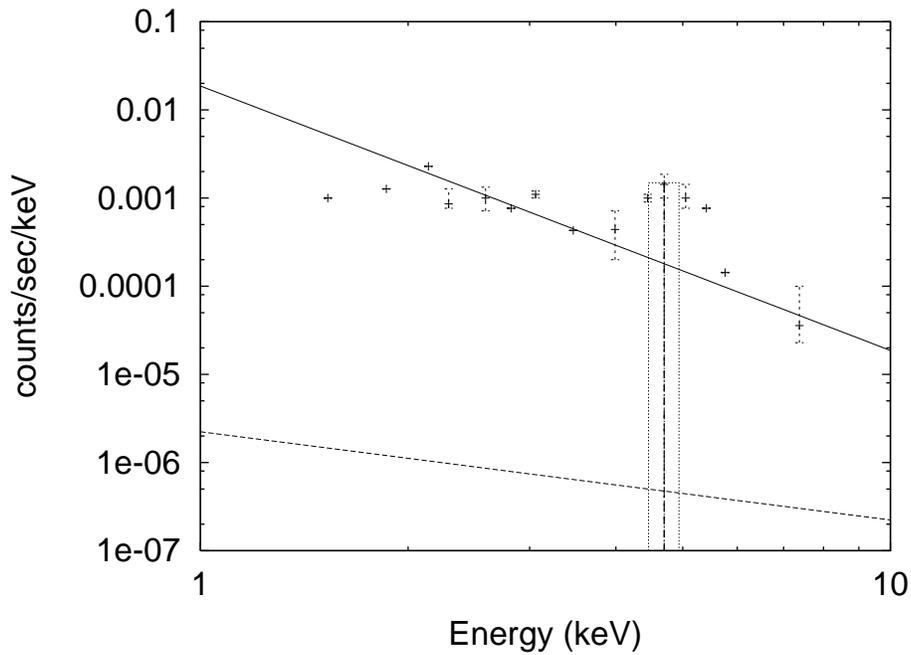}
 \caption{ Integrated spectrum:solid line, afterglow emission toward the observer; dashed line (vertical), fluorescent line emission of Fe K$\alpha$ originated from the ring  illuminated by the X-ray afterglow in the energy bin of 0.48 keV; dotted line, thermal Bremsstrahlung emission from the ring shocked by the incoming fireball; crosses, the observation of the afterglow and the Fe line \cite{rf:Antonelli}}
 \label{Fig:2}
\end{center}
\end{figure}

\end{document}